\documentclass{IEEEtran}

\usepackage{graphicx}
\usepackage{lineno}
\modulolinenumbers[5]
\usepackage[cmex10]{amsmath}
\usepackage{amsfonts}

\usepackage{color}
\usepackage{algorithm}
\usepackage{algorithmic}
\usepackage{array}
\usepackage{multirow}
\usepackage{booktabs}
\usepackage{subfigure}
\usepackage{bm}

\allowdisplaybreaks[4]
\usepackage{underscore}
\usepackage{stfloats}
\usepackage{float}
\usepackage{lipsum}
\usepackage{cite}
\usepackage{amssymb}
\usepackage{amsmath}
\usepackage{mathrsfs}
\usepackage[margin=0.75in]{geometry}
\usepackage{amsthm}

\makeatletter
\begin{document}

\title{
  A Theoretically-Grounded Codebook for Digital Semantic Communications 

  % \author{Lingyi Wang
  % \thanks{Lingyi Wang is with the Bradley Department of Electrical and Computer Engineering, Virginia Tech, Alexandria, VA, 22305, USA
  % (e-mail: lingyiwang@vt.edu).}
  % }

\author{
\IEEEauthorblockN{
Lingyi Wang\IEEEauthorrefmark{1}, 
Rashed Shelim\IEEEauthorrefmark{1},  
Walid Saad\IEEEauthorrefmark{1},
and Naren Ramakrishnan\IEEEauthorrefmark{2}\\
\IEEEauthorblockA{\IEEEauthorrefmark{1}Bradley Department of Electrical and Computer Engineering, Virginia Tech, Alexandria, VA, 22305, USA}\\
\IEEEauthorblockA{\IEEEauthorrefmark{2}Department of Computer Science, Virginia Tech, Alexandria, VA, 22305, USA}\\
\IEEEauthorblockA{Emails: \{lingyiwang, rasheds, walids, naren\}@vt.edu
}}
}
}
\thispagestyle{empty}
\maketitle
\thispagestyle{empty}

\begin{abstract} 
 The use of a learnable codebook provides an efficient way for semantic communications to map vector-based high-dimensional semantic features onto discrete symbol representations required in digital communication systems. In this paper, the problem of codebook-enabled quantization mapping for digital semantic communications is studied from the perspective of information theory. Particularly, a novel theoretically-grounded codebook design is proposed for jointly optimizing quantization efficiency, transmission efficiency, and robust performance. First, a formal equivalence is established between the one-to-many synonymous mapping defined in semantic information theory and the many-to-one quantization mapping based on the codebook's Voronoi partitions. Then, the mutual information between semantic features and their quantized indices is derived in order to maximize semantic information carried by discrete indices. To realize the semantic maximum in practice, an entropy-regularized quantization loss based on empirical estimation is introduced for end-to-end codebook training. Next, the physical channel-induced semantic distortion and the optimal codebook size for semantic communications are characterized under bit-flip errors and semantic distortion. To mitigate the semantic distortion caused by physical channel noise, a novel channel-aware semantic distortion loss is proposed. Simulation results on image reconstruction tasks demonstrate the superior performance of the proposed theoretically-grounded codebook that achieves a 24.1\% improvement in peak signal-to-noise ratio (PSNR) and a 46.5\% improvement in learned perceptual image patch similarity (LPIPS) compared to the existing codebook designs when the signal-to-noise ratio (SNR) is 10 dB.
\end{abstract}

\begin{IEEEkeywords}
  Digital semantic communication systems, codebook-enabled vector quantization, information theory. 
\end{IEEEkeywords}
  
\IEEEpeerreviewmaketitle
  
\section{Introduction}
Semantic communication is emerging as a promising communication paradigm that shifts the focus from conventional bit-accurate delivery to goal-driven, meaning-preserving transmission \cite{chaccour2022less,10929033,wang2023adaptive}. Particularly, in an end-to-end semantic communication framework, the transmitter typically uses deep neural networks to extract semantic features from raw data based on communication goals \cite{wang2023adaptive}.  However, existing digital baseband hardware and protocol stacks support only discrete bitstreams for channel coding and modulation \cite{10101778}.  Hence, it is challenging for semantic communication systems to efficiently and reliably convert the high-dimensional continuous semantic features into discrete indices with preserved critical meaning \cite{10960697}. To address this challenge, learnable codebooks for vector quantization can be applied to discretize the semantic features without losing semantic capacity \cite{ye2025low,10654371,10622764}. Specifically, the use of a codebook allows the semantic communication system to partition the semantic feature space into a collection of disjoint subspaces and label these subspaces with a finite integer index set. Then, each semantic feature vector can be quantized to a discrete index by finding its optimally associated subspace. 

Recently, a number of works studied the problem of codebook-enabled semantic communications \cite{10101778,10960697,ye2025low,10654371,10622764,10978541,10431795,zhou2024moc,shin2025esc}. Several prior works in \cite{10101778,10960697,ye2025low,10654371} primarily used a codebook for compact semantic discretization to reduce communication costs by quantization loss-based end-to-end training. In \cite{10622764} and \cite{10431795}, the authors proposed unified codebooks for multi-task semantic communications by exploiting semantic relationships among tasks. The work in \cite{zhou2024moc} investigated a multilevel codebook that used multi-head octonary quantization to further compress indices. The authors in \cite{shin2025esc} proposed the multi-codebook design with each codebook trained with different bit-flip probabilities for adaptive modulation.
However, despite its success in addressing some meaningful problems, this prior work \cite{10101778,10960697,ye2025low,10654371,10622764,10978541,10431795,zhou2024moc,shin2025esc} is limited in a number of ways:
\begin{itemize}
  \item There is a lack of a unified theoretical foundation on existing codebooks for semantic communications. Moreover, it is hard to determine the theoretically optimal codebook size.
  \item Existing codebook schemes solely optimize the quantization distortion, which leads to codeword under-utilization or over-concentration.
  \item Existing codebook schemes cannot explicitly model the impact of physical-layer bit-flip errors on semantic representations, thus lacking channel-aware distortion metrics along with robustness optimization.
\end{itemize}
 
The main contribution of this paper is to overcome the aforementioned limitations of existing codebooks for digital semantic communications by rethinking the codebook-enabled vector quantization from a perspective of information theory.
In particular, we propose a novel theoretically-grounded codebook design for better quantization efficiency, transmission efficiency, and robust performance. First, we demonstrate that the critical one-to-many synonymous mapping defined in semantic information theory can be realized through the many-to-one quantization mapping over the codebook's Voronoi partitions. Our perspective provides an unified theoretical-engineering model for codebook-enabled semantic communications. Then, we derive the mutual information between semantic features and their quantized indices. To maximize the semantic information carried by discrete indices, we propose an entropy-regularized quantization loss with empirical estimation to ensure balanced codeword utilization. Moreover, we characterize the channel-induced semantic distortion under bit-flip errors and the optimal distortion-aware codebook size.
Driven by the channel-induced semantic distortion, we propose a novel channel-aware distortion loss to address semantic distortion induced by bit-flip errors. Extensive simulations on an image reconstruction task demonstrate that the proposed theoretically-grounded codebook achieves a 24.1\% improvement in peak signal-to-noise ratio (PSNR) and a 46.5\% improvement in learned perceptual image patch similarity (LPIPS) compared to the existing codebook designs when the signal-to-noise ratio (SNR) is 10 dB.

The rest of this paper is organized as follows. In Section II, we present the codebook-enabled digital semantic communications. Then, we propose a novel theoretically-grounded codebook in Section III. In Section IV, we demonstrate the simulation results and analysis.
Finally, conclusions are drawn in Section V. 

\begin{figure*}[t!]
  \centering
  \includegraphics[width=0.9\linewidth]{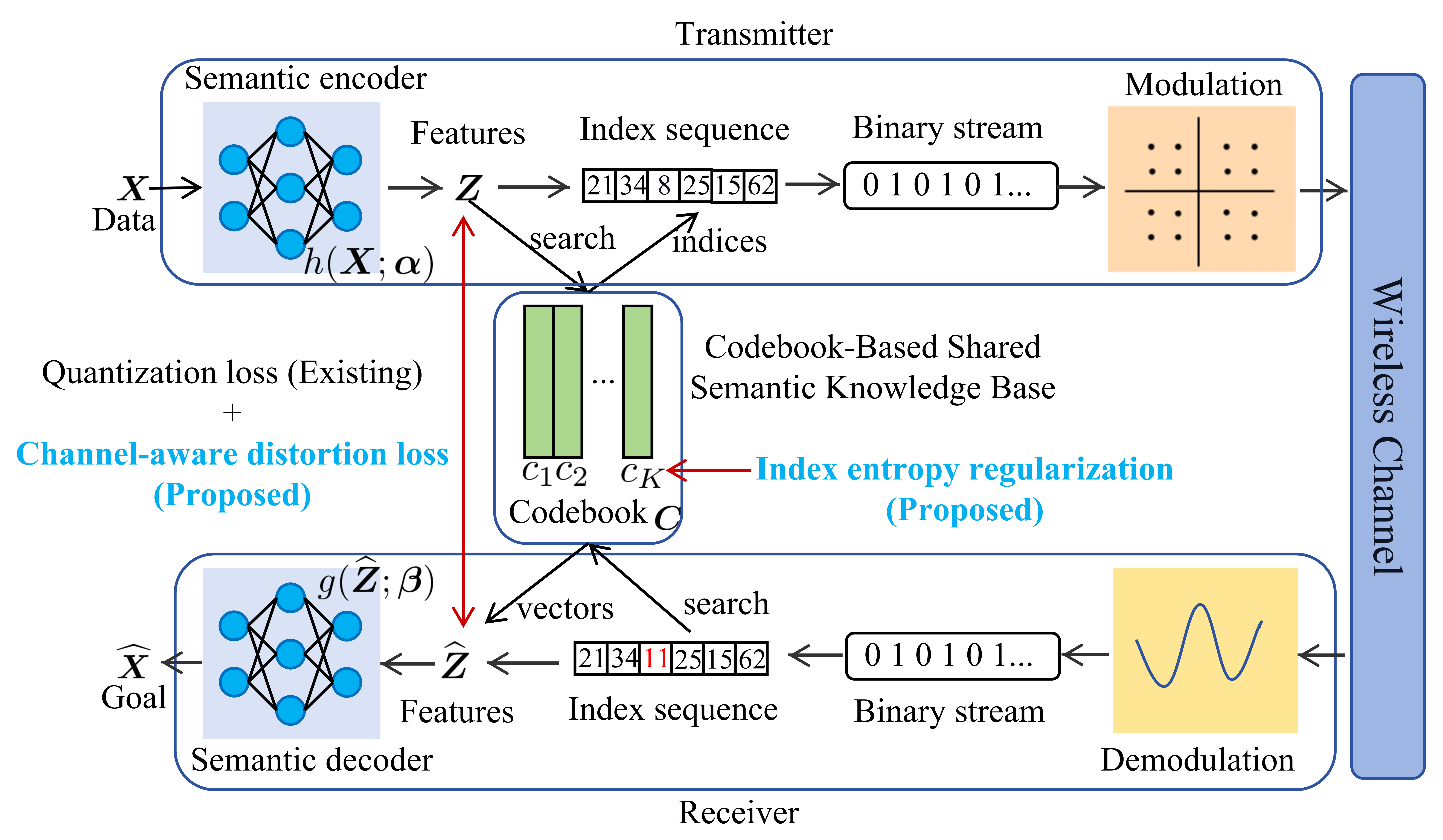}
  \vspace{-0.2cm}
  \caption{Illustration of a codebook-enabled digital semantic communication framework.}
  \vspace{-0.4cm}
\end{figure*}

\section{Codebook for Digital Semantic Communications}
As shown in Fig.~1, we consider a codebook-enabled digital semantic communication network, in which, without loss of generality, the source data consists of images. Let the image dataset be $\boldsymbol{\mathcal{X}}$, with each image $\boldsymbol{X} \in \boldsymbol{\mathcal{X}} \subset \mathbb{R}^{H\times W \times O}$, where $H$, $W$ and $O$ are the height, width, and the channel size of each input image, respectively. The learnable codebook for vector quantization is given by
$\boldsymbol{C} \triangleq \left[\boldsymbol{c}_{1},\boldsymbol{c}_{2},\dots,\boldsymbol{c}_{k},\dots,\boldsymbol{c}_{K}\right]\subset\mathbb{R}^N$,
where $K$ is the codebook size, and each element \(\boldsymbol{c}_k \in \mathbb{R}^{N}\) is called a \emph{semantic codeword} with dimension size $N$ and index $k$. The codebook $\boldsymbol{C}$ serves as a shared, constant knowledge base for the transmitter and the receiver.
  
At the transmitter, the semantic encoder $h(\boldsymbol{X};\alpha)$ with parameters $\alpha$ extracts semantic features $\boldsymbol{Z} \in \mathbb{R}^{M \times N}$ from the raw image $\boldsymbol{X}$.
The learnable codebook $\boldsymbol{C}$ then serves to map vector-based semantic representations onto discrete symbol representations in the digital communication system.  Particularly, each semantic vector $\boldsymbol{z}_{m} \in \boldsymbol{Z} \subset \mathbb{R}^{N}$ with index $m \leq M$ is quantized by finding its nearest semantic codeword $\boldsymbol{c}_k\in \boldsymbol{C}$. The \emph{quantization mapping} $q\left(\cdot\right)$ by the nearest-neighbor algorithm in the codebook-enabled semantic space will be
\begin{equation}
q\left(\boldsymbol{z}_{m}\right)
\;=\;
\arg\min_{k\in\{1,\dots,K\}}
\left\|\boldsymbol{z}_{m}-\boldsymbol{c}_k\right\|_2.
\end{equation} 
Then, with each vector $\boldsymbol{z}_{m} \in \boldsymbol{Z}$ mapped to an index by (1), the index sequence of semantic features $\boldsymbol{Z}$ is obtained by
\begin{equation}
\boldsymbol{S} = \left[s_{m}\right]_{m=1}^M
\;\in\;\{1,\dots,K\}^{M},
\quad
s_{m} = q(\boldsymbol{z}_{m}).
\end{equation}
The codebook-enabled vector quantization discretizes the original floating-point semantics into compact integer indices, which stand in for the continuous feature.  
The resulting index sequence is then processed by standard digital communication blocks, including channel coding, mapping of constellation symbols, modulation over carriers, and transmission over the physical channel.

The preimage of mapping \(q\) is represented by
\begin{equation}
  \begin{aligned}
    &q^{-1}(k)=\left\{\boldsymbol{z}\in\mathbb{R}^N : q(\boldsymbol{z})=k\right\} \\
    &\quad \, = \left\{\,\boldsymbol{z}\in\mathbb{R}^N : \|\,\boldsymbol{z} - \boldsymbol{c}_k\|_2 \le \|\,\boldsymbol{z} - \boldsymbol{c}_j\|_2,\;\forall\,j\neq k \right\},
  \end{aligned}
\end{equation}
which is a Voronoi region of \(\mathbb{R}^N\) including infinite distinct semantic vectors. This Voronoi region can be considered as a \emph{semantic feature cluster}, which can be obtained by
\begin{equation}
\mathbb{R}^N
\;=\;
\bigcup_{k=1}^K
q^{-1}(k),
\quad
q^{-1}(k)\,\cap\,q^{-1}(\ell)=\varnothing
\ \ (k\neq\ell).
\end{equation}
(3) and (4) partition the semantic feature space into disjoint Voronoi regions, thus enabling quantization by mapping each feature vector to a unique representative codeword.
Hence, the codebook induces an equivalence-class partition \(\boldsymbol{Z}/q=\{q^{-1}(1),\dots,q^{-1}(K)\}\) of the continuous semantic space.   
The receiver demodulates and channel-decodes the received symbols to recover the integer index sequence $\widehat{\boldsymbol{S}}=\left[\widehat{s}_{m}\right]_{m=1}^M$ and recovers the semantic information $\widehat{\boldsymbol{z}}_{m} = \boldsymbol{c}_{\widehat{s}_{m}} = q^{-1}(\widehat{s}_{m})$.
The recovered semantic features $\widehat{\boldsymbol{Z}}\in \mathbb{R}^{M \times N}$ are input into the semantic decoder $g(\cdot;\beta)$ with parameters $\beta$ to accomplish the semantic task by $\widehat{\boldsymbol{X}} = g(\widehat{\boldsymbol{Z}};\beta)$. To train the learnable codebook, the quantization loss \cite{10101778,10960697,ye2025low,10654371,10622764,10978541,10431795,zhou2024moc,shin2025esc} is minimized by 
\begin{equation}
  \mathcal{L}^{\mathrm{qua}}
  =\mathbb{E}_{\boldsymbol{Z}\sim p_{\boldsymbol{Z}}}\left[\| \sum \boldsymbol{z} -\boldsymbol{c}_{q(\boldsymbol{z})}\|^2\right].
\end{equation}

Although (5) can measure the semantic distortion caused by the quantization mapping, it cannot ensure the maximum semantic information carried by each index, s.e., efficient semantic codeword design, and explicitly evaluate the semantic distortion from the physical channel noise. 

\section{Theoretically-Grounded Codebook}
\begin{figure}[t!]
  \centering
  \includegraphics[width=1\linewidth]{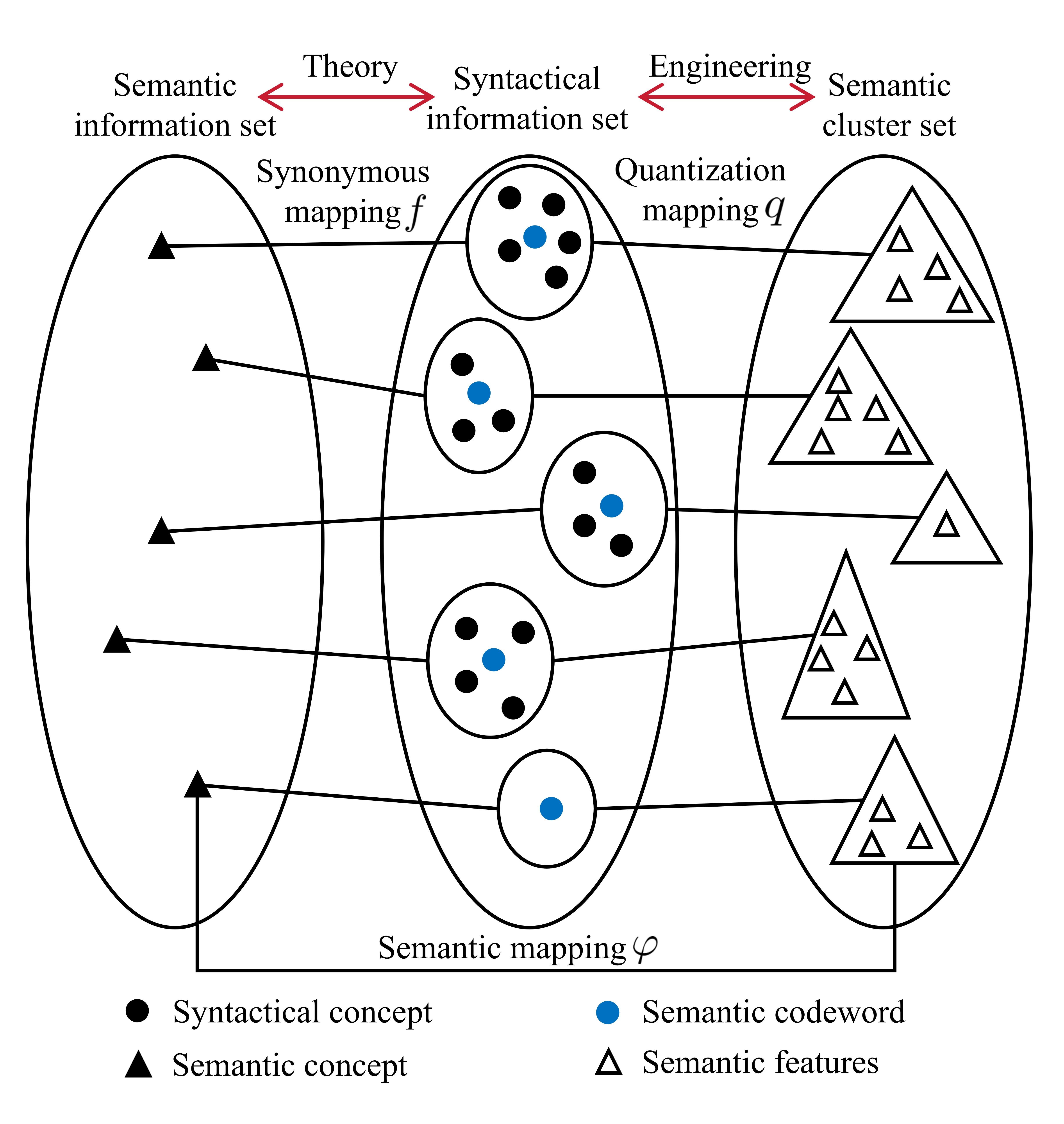}
  \vspace{-0.7cm}
  \caption{A showcase of the unified synonymous mapping and quantization mapping from a joint perspective of theory and engineering.}
  \label{fig:2}
\end{figure}

In this section, we first rethink the quantization mapping of the codebook from a perspective of semantic information theory, as shown in Fig. \ref{fig:2}. Then, a theoretically-grounded codebook is proposed with the entropy regularization for the maximum mutual information between semantic features and their quantized discrete indices, along with a novel channel-aware semantic distortion loss to alleviate the impact of physical channel noise.

\subsection{Relationships Between Synonymous Mapping And Quantization Mapping}
Semantic information theory \cite{niu2024mathematical}
is based on a consensus that there is a one-to-many synonymous mapping $f$ between the semantic source $\widetilde{U}$ and the syntactic source $U$ to generate the raw message, as shown in Fig. \ref{fig:2}. For instance, the semantic concept of ``happiness'' can be represented by various syntactic realizations such as text information ``joy'' and smiling images.  Let the syntactic set be $U = \{u_1,\dots,u_j,\dots,u_J\}$ and the semantic set be
$\widetilde U = \{\widetilde u_1,\dots,\widetilde u_{\widetilde{j}},\dots,\widetilde u_{\widetilde{J}}\}$, $\widetilde{J} \ll J$, where $u_j$ represents a syntactic realization and $\widetilde u_{\widetilde{j}}$ represents a semantic concept.
Then, the synonymous mapping is given by $f:\,\widetilde U \to U$. The synonymous mapping $f$ induces an equivalence-class partition partition as
$U/f=\{\,f(\widetilde u_1),\dots,f(\widetilde u_{\widetilde{j}}),\dots,f(\widetilde u_{\widetilde{J}})\}$, which can be obtained by
\begin{equation}
U = \bigcup_{\widetilde{j}=1}^{\widetilde J} f(\widetilde u_{\widetilde{j}}), \quad U_f(\widetilde u_{\widetilde{j}})\cap f(\widetilde u_{\widetilde{m}})=\varnothing \ \ (\widetilde u_{\widetilde{j}} \neq \widetilde u_{\widetilde{m}}).
\end{equation}

In practice, the learnable codebook realizes the synonymous mapping $f$ by the quantization mapping $q$. In particular, as discussed in (1)--(4), the learnable codebook $\boldsymbol{C} \triangleq \left[\boldsymbol{c}_{1},\boldsymbol{c}_{2},\dots,\boldsymbol{c}_{k},\dots,\boldsymbol{c}_{K}\right]\subset\mathbb{R}^N$, partitions the continuous semantic feature space $\mathbb{R}^N$ into $K$ Voronoi cells, where each semantic codeword $\boldsymbol{c}_k$ provides a \emph{standard syntax} for a set of synonymous syntaxes $f(\widetilde u_{\widetilde{j}})$.
Since the semantic information is embedded in and cannot be directly separated from its syntactic representations, the semantic encoder is applied to infer a latent meaning representation with semantic features $\boldsymbol{z}$ from the observed data $\boldsymbol{X}$. 
However, in practice, the same semantic concept can produce a semantic feature cluster, represented by $q^{-1}(k)$, due to morphological variability and observation noise. This reflects the fact that a single meaning can manifest through diverse syntactic expressions. 
During quantization mapping, regardless of its low-level syntactic variations, any semantic feature vector in the Voronoi region of $\boldsymbol{c}_k$ is assigned the same integer index $k$.  In this way, the codebook-enabled vector quantization acts as the many-to-one mapping from syntactic expressions to a single semantic symbol. Let the one-to-many semantic mapping between semantic concept $\widetilde u$ and semantic features $\boldsymbol{z}$ be $\varphi: {\widetilde{U}} \to \mathbb{R}^{N}$. The semantic mapping $\varphi$ means that one semantic concept can behave different features with different contexts, and coding noise from source data. 
The \(q\circ\varphi\) and \(f\) coincide on every semantic concept as
\begin{equation}
f(\tilde u)=U_k
\;\Longleftrightarrow\;
\varphi(\widetilde u)\in q^{-1}(k)
\;\Longleftrightarrow\;
q\left(\varphi(\widetilde u)\right)=k.
\end{equation}
Hence, from the perspective of information theory, the codebook can preserve the theoretical advantages of synonymic diversity with Voronoi regions partitioned by the quantization mapping, and enhances the transmission efficiency by providing discrete semantic representations that can be directly deployed in existing digital communication systems.

\subsection{Entropy Regularized Indices}  
Now, we derive the entropy regularization based on mutual information between semantic features and their quantized indices to maximize semantic information carried by discrete indices.
Let $p_{\boldsymbol{Z}}(\boldsymbol{z})$ be the density of the continuous semantic feature vector \(\boldsymbol{z}\in\mathbb{R}^N\), and then we have $p_{\boldsymbol{Z},\boldsymbol{S}}(\boldsymbol{z},k) = p_{\boldsymbol{Z}}(\boldsymbol{z})\,\mathbb{I}\{q(\boldsymbol{z})=k\}$, where \(\mathbb{I}(x)\) is a binary-valued indicator function that equals to 1 if the condition \( x \) holds true and 0 otherwise.
By definition of information theory, the mutual information between semantic features and their quantized indices is
\begin{equation}
\begin{aligned}
s(\boldsymbol{Z};\boldsymbol{S})
&=\sum_{k=1}^{K}\int p_{\boldsymbol{Z},\boldsymbol{S}}(\boldsymbol{z},k)\,\log\frac{p_{\boldsymbol{Z},\boldsymbol{S}}(\boldsymbol{z},k)}{p_{\boldsymbol{Z}}(\boldsymbol{z})\,p_{\boldsymbol{S}}(k)}\,dz\\
&=H(\boldsymbol{S})-H(\boldsymbol{S}\mid \boldsymbol{Z})\,=\,H(\boldsymbol{S}),
\end{aligned}
\end{equation}
where \(H(\boldsymbol{S}\mid \boldsymbol{Z})=0\) since the index sequence \(\boldsymbol{S}\) is determined given semantic features \(\boldsymbol{Z}\) by the many-to-one quantization mapping $q$.
Let the occurrence probability of each index be \(\pi_k =\Pr[q(\boldsymbol{z})=k]\). Equivalently, the coding-rate can be rewritten as
\begin{equation}
  \begin{aligned}
&s(\boldsymbol{Z};\boldsymbol{S})
=\int p_{\boldsymbol{Z}}(\boldsymbol{z})\,\log\frac{p(s=q(\boldsymbol{z})\mid \boldsymbol{z})}{\pi_{q(\boldsymbol{z})}}\,d\boldsymbol{z} \\
&=\int p_{\boldsymbol{Z}}(\boldsymbol{z})\,\log\frac{1}{\pi_{q(\boldsymbol{z})}}\,d\boldsymbol{z}
=-\mathbb{E}\left[\log\pi_{\boldsymbol{S}}\right] \leq \log K,
  \end{aligned}
\end{equation}
where the equality holds when $\pi_1=\pi_2= \cdots = \pi_K = 1/K$.
Hence, to increase the effective semantic information carried by the discrete indices and improve the transmission efficiency, we need to ensure every codeword is equitably accessible, which can be realized by introducing the entropy regularization of indices. However, in practice, the actual distribution $p_{\boldsymbol{Z}}(\boldsymbol{z})$ is unknown, and, thus, it is hard to directly calculate the entropy $H(\boldsymbol{S})$ for end-to-end codebook training. 
Here, we utilize the batch samples $\boldsymbol{Z} \sim p_{\boldsymbol{Z}}$ to do empirical estimation, and the estimated entropy $\widehat H(\boldsymbol{S})$ of indices can be represented by
\begin{equation}
\widehat H(\boldsymbol{S})
=-\sum_{k=1}^K \widehat\pi_k\log\widehat\pi_k,
\end{equation}
with
\begin{equation}
  \widehat\pi_k
=\mathbb{E}_{\boldsymbol{Z}\sim p_{\boldsymbol{Z}}}\left[\sum_{z \in \boldsymbol{Z}}\mathbb{I}\{q(\boldsymbol{z})=k\}\right].
\end{equation}
Then, the estimated mutual information between semantic features and their quantized indices can be rewritten as
\begin{equation}
\widehat s(\boldsymbol{Z};\boldsymbol{S})=\widehat H(\boldsymbol{S})
=-\mathbb{E}_{\boldsymbol{Z}\sim p_{\boldsymbol{Z}}}\left[\sum\log\widehat\pi_{q(\boldsymbol{z})}\right].
\end{equation}
Hence, the quantization loss with entropy regularization for end-to-end codebook training can be represented by
\begin{equation}
\mathcal{L}^{\mathrm{reg}}
=\mathbb{E}_{\boldsymbol{Z}\sim p_{\boldsymbol{Z}}}\left[\| \sum \boldsymbol{z} -\boldsymbol{c}_{q(\boldsymbol{z})}\|^2\right]
-\gamma\,\widehat H(\boldsymbol{S}),
\end{equation}
where \(\gamma>0\) is a parameter that captures the tradeoff between quantization fidelity and index-entropy.  The gradient with respect to the empirical frequencies \(\widehat\pi_k\) is
\begin{equation}
  \frac{\partial(-\widehat H(\boldsymbol{S}))}{\partial\widehat\pi_k}
=1+\log\widehat\pi_k,
\end{equation}
which drives the occurrence probability \(\widehat\pi_k\) of each index toward the uniform distribution \(1/K\).

\subsection{Physical Channel-Induced Semantic Distortion}
Now, we explore the semantic distortion over the physical channel. In particular, each index \(s \in\{1,\dots,K\}\) is encoded into a \(L\)-length binary sequence \(b\in\{0,1\}^L\), where \(L=\left\lceil \log_2 K \right\rceil\), and $\lceil x \rceil$ returns the smallest integer $\geq x$. We consider the indices are transmitted over a memoryless binary symmetric channel with bit-flip probability \(p\).  Let \(\widehat b\) be the received binary vector, and \(d_H(b,\widehat b)\) be Hamming distance.  Then, the error-weight distribution of the binary index sequence can be obtained by
\begin{equation}
\Pr\left(d_H(b,\widehat b)=w\right)
=\binom{L}{w}\,p^w\,(1-p)^{L-w},
w \leq L,
\end{equation}
and the overall index-error probability can be represented by
\begin{equation}
P_e
=\Pr\left(\widehat b\neq b\right)
=1-\Pr\left(d_H(b,\widehat b)=0\right)
=1-(1-p)^L.
\end{equation}
The conditional probability with random labeling can be approximately represented by
\begin{equation}
\Pr(\,\widehat{s} = \ell \mid s = k)
\;=\;
\begin{cases}
1 - P_e, & \ell = k,\\
\dfrac{P_e}{K-1}, & \ell \neq k.
\end{cases}
\end{equation}
Then, the physical channel-induced semantic distortion based on average squared-error is obtained by
\begin{equation}
\begin{aligned}
D_{\mathrm{ch}}&=
\sum_{k=1}^K \pi_k\left[
   (1-P_e)\cdot 0
   + \sum_{\ell\neq k}\frac{P_e}{K-1}\|\boldsymbol{c}_k - \boldsymbol{c}_\ell\|_2^2
\right] \\
& = P_e \sum_{k=1}^K \pi_k\,\bar\Delta_k^2,
\end{aligned}
\end{equation}
with
\begin{equation}
\bar{\Delta}^2_k
=\frac{1}{K-1}\sum_{k\neq\ell}\|\boldsymbol{c}_k - \boldsymbol{c}_\ell\|_2^2.
\end{equation}
Thus, the total semantic distortion induced by the physical channel noise can be expressed as
\begin{equation}
D_{S}(K,p)
=\underbrace{\mathbb{E}_{\boldsymbol{Z} \sim p_{\boldsymbol{Z}}}\left[\| \sum \boldsymbol{z} -\boldsymbol{c}_{q(\boldsymbol{z})}\|^2\right]}_{\text{quantization loss}} + D_{\mathrm{ch}}(K,p).
\end{equation}
In traditional communication systems, a codebook is designed to map fixed-length bit sequences to modulation symbols, typically by maximizing minimum Euclidean distance or optimizing channel capacity under bit-error probability $p$. However, in the semantic communication framework, the codebook is constructed to quantize high-dimensional features to accurately convey the semantic information and directly minimize semantic distortion $D_S(K, p)$ rather than bit-level errors. In this context, the optimal codebook size $K^*$ is therefore chosen to balance semantic fidelity against bandwidth cost. Hence,
given the bit-error probability $p$ and the bandwidth weighting factor $\lambda$, the optimal codebook size for semantic communications can be selected by
\begin{equation}
K^*
=\arg\min_{K}\left\{D_{S}(K,p)\;+\;\lambda\,R(K)\right\},
\end{equation}
where \(R(K)=M\log_2K\) is the total bit-rate. 

In addition to the selection of the optimal codebook size, the conventional codebooks \cite{10101778,10960697,ye2025low,10654371} for semantic communications only minimize the quantization error and cannot capture the impact of physical-layer bit flips on digital communication. Thus, existing semantic codebooks experience severe semantic distortion or even total information loss under high error-rate conditions.
Driven by channel-induced semantic distortion (13), we further introduce the generalized channel-aware distortion loss, which can be represented by
\begin{equation}
  \mathcal{L}^{\mathrm{ch}} = \mathbb{E}_p\left[D_{\mathrm{ch}}(K,p)\right] = \mathbb{E}_p\left[P_e \sum_{k=1}^K \pi_k\,\bar\Delta_k^2\right],
\end{equation}
which imposes structured constraints on the codebook to enhance the robustness of semantic communications. The total loss function for the codebook can be represented by
\begin{equation}
  \mathcal{L}^{\boldsymbol{C}}
  =\mathbb{E}_{\boldsymbol{Z}\sim p_{\boldsymbol{Z}}}\left[\| \sum \boldsymbol{z} -\boldsymbol{c}_{q(\boldsymbol{z})}\|^2\right] + \omega \mathcal{L}_{\mathrm{ch}}
  -\gamma\,\widehat H(\boldsymbol{S}),
\end{equation}
where $\omega \in (0,1)$ is the weight factor for the channel-aware semantic distortion.

Fundamentally, the proposed entropy regularization and the channel-aware semantic distortion loss enable the codebook to find an optimal space partition scheme. In particular, the entropy regularization of discrete indices encourages more fair partitions of the semantic feature space based on the occurrence probability of each index, thus maximizing the semantic information carried by each index. The channel-induced distortion loss reduces the spatial distance between the original features and distorted quantization based on the possibility of physical-layer bit-flip errors, which alleviates semantic distortion and enhances semantic robustness.

\section{Simulation Results And Analysis}
  \subsection{Simulation Setup}
  For our simulations, an image semantic communication network with the reconstruction task is considered with a single-antenna transmitter and a single-antenna receiver, where the semantic encoder and decoder are based on vector quantized-variational autoencoder (VQ-VAE) \cite{van} with the same training parameters provided in \cite{10622764}. The end-to-end training loss of the semantic network for image reconstruction can be represented by
\begin{equation}
  \begin{aligned}
    \mathcal{L}(\alpha, \beta, \boldsymbol{C}) \,=  \,
    \underbrace{\left\|G_{\ell}^j(\boldsymbol{X})-G_{\ell}^j(\widehat{\boldsymbol{X}})\right\|_2^2}_{\text{ reconstruction loss}} \, + \, \mathcal{L}^{\boldsymbol{C}}.
  \end{aligned}
\end{equation}
The reconstruction loss is based on the pretrained VGG16 network \cite{simonyan2014very} to capture the semantic similarity in the latent space, where $G_{\ell}^j$ is the Gramm matrix, represented by 
\begin{equation}
  G_j^\ell(\boldsymbol{X})=\frac{1}{H_j W_j} \sum_{h=1}^{H_j} \sum_{w=1}^{W_j} \ell_j(\boldsymbol{X})_{o,h,w} \ell_j(\boldsymbol{X})_{o^{\prime},h,w},
\end{equation}
  and $\ell_j(\cdot)$ represents the first $j$ layers of VGG16 with parameter $\ell$, $o\leq O_j$. We set $j=8$, $\omega=0.1$ and $\gamma=0.1$.
  The modulation scheme is set to 64-QAM \cite{zhou2024moc}, and the codebook size is set to $K=256$.
  We adopt VOC-2012 as the training database, which consists of $11,530$ images of $20$ classes. We introduce the JPEG+LDPC scheme based on joint photographic experts group (JPEG) and 1/2-rate low-density parity-check (LDPC) code, the BPG+LDPC scheme based on better portable graphics (BPG) and the 1/2-rate LDPC code, and the VQ-VAE scheme \cite{10622764} that only considers quantization loss as the benchmark schemes. LPIPS \cite{zhang2018unreasonable} and PSNR are used to evaluate the quality of reconstructed images.

 Fig. 3 and Fig. 4 respectively show the LPIPS performance and the PSNR performance of different schemes across different SNRs in a Rayleigh fading channel.
 Compared to the existing VQ-VAE-based scheme, the proposed theoretically-grounded codebook achieves a 24.1\% improvement in PSNR and a 46.5\% improvement in LPIPS compared to the existing VQ-VAE method when SNR is 10 dB. These improvements can be attributed to the proposed index-entropy regularization and channel-aware semantic distortion loss, which is further demonstrated by the ablation experiments. In particular, we independently introduce the regularization item and channel-aware distortion loss to the VQ-VAE-based schemes, respectively named ``VQ-VAE + Index Entropy'' and ``VQ-VAE + Channel-Aware''. 
 Under low-SNR conditions with high bit-flip rates, the quantized indices of the semantic features can be distorted over the physical channel.
In this context, the channel-aware semantic distortion loss can reduce the spatial distance between the original features from the transmitter and the distorted features at the receiver based on distortion probability. The distortion loss suppresses the semantic distortion caused by error bits, thus enhancing the semantic robustness. 
In the context of high SNRs with accurate symbol transmission, the entropy regularization of discrete indices encourages more fair partitions of the semantic feature space based on the occurrence probability of each index, and maximizes the information carried by each index, thus improving transmission efficiency. Hence, the channel-aware semantic distortion loss and entropy regularization can enable an efficient codebook design by the optimal space partitions.

  \begin{figure}[t]
    \centering
    \includegraphics[width=1\linewidth]{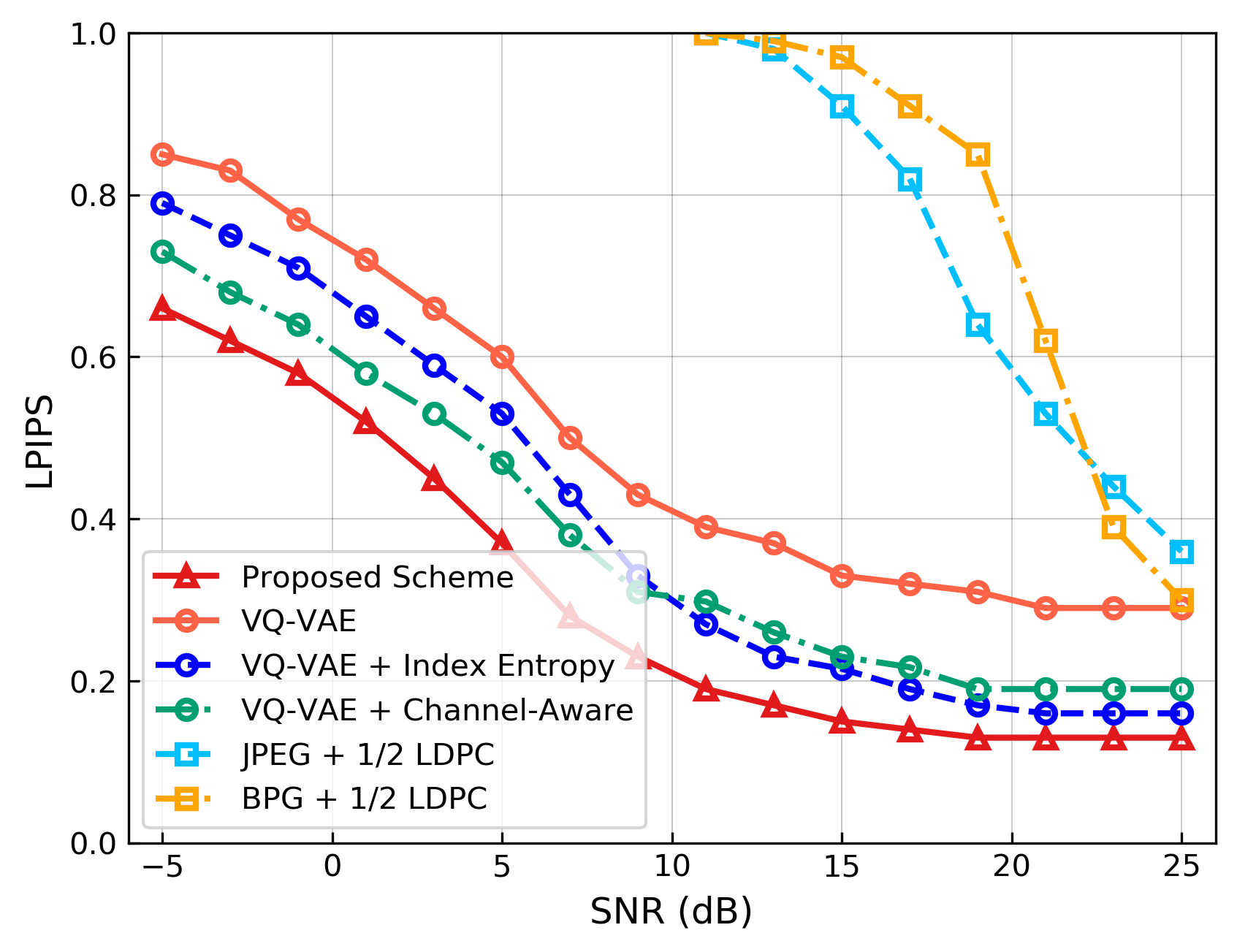}
    \vspace{-0.6cm}
    \caption{The LPIPS performance of different schemes across different SNRs.}
    \vspace{-0.35cm}
  \end{figure}

\section{Conclusion}
In this paper, we have rethought the codebook-enabled vector quantization for digital semantic communications through an information-theoretic perspective and introduced a novel, theoretically-grounded codebook design that jointly enhances quantization efficiency, transmission efficiency, and robustness. We have demonstrated that the one-to-many synonymous mapping of semantic information theory can be realized through the Voronoi regions partitioned by the quantization mapping, thus unifying abstracted semantic information theory with the codebook-based implementation. Building on this foundation, we have derived the mutual information between high-dimensional semantic features and their quantized discrete indices, and have introduced the index-entropy regularization. The entropy-regularized quantization loss has encouraged the balanced codeword utilization and maximized the semantic information carried by each index. We have further analyzed the impact of physical-layer bit-flip errors on semantic distortion, proposed the channel-aware semantic distortion loss for robust performance, and presented the optimal codebook size under error-rate and bandwidth constraints.
Extensive simulations on image reconstruction tasks have demonstrated that the proposed theoretically-grounded codebook achieves 24.1\% improvement in PSNR and 46.5\% improvement in LPIPS compared to the existing VQ-VAE designs when SNR is 10 dB.

\begin{figure}[t]
  \centering
  \includegraphics[width=1\linewidth]{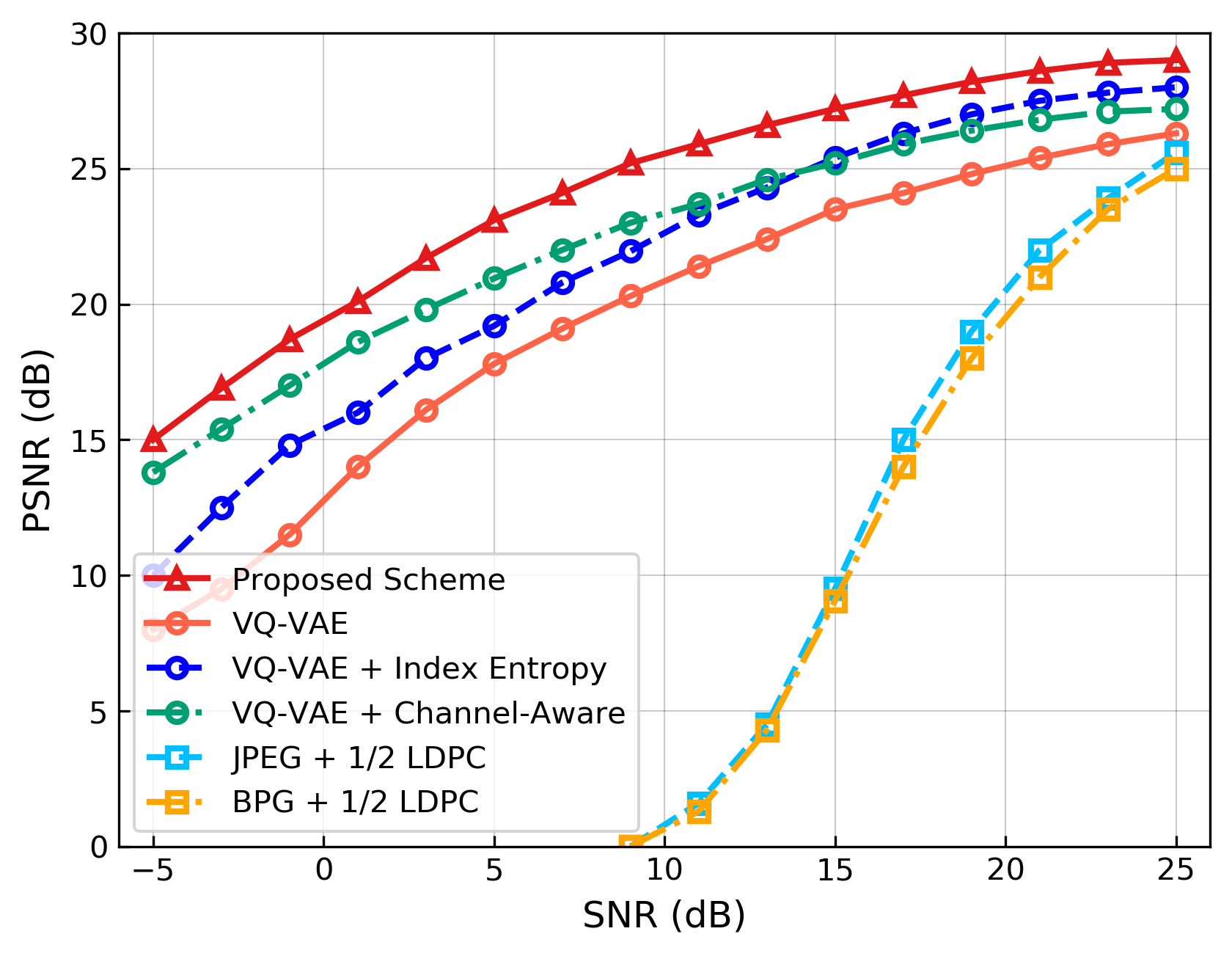}
  \vspace{-0.66cm}
  \caption{The PSNR performance of different schemes across different SNRs.}
  \vspace{-0.4cm}
\end{figure}

\bibliography{IEEEabrv,reference}

% that's all folks
\end{document}